\documentclass[apl]{revtex4}%

\usepackage{graphicx}
\usepackage{dcolumn}
\usepackage{bm}
\usepackage{makecell} 
\usepackage[english]{babel}
\usepackage{verbatim}
\begin{document}
\preprint{}

\title{Effect of structural disorder on quantum oscillations in graphite}

\author{B. C. Camargo}
 \email{b.c_camargo@yahoo.com.br}
\author{Y. Kopelevich}%
\affiliation{%
Instituto de Fisica “Gleb Wataghin”, Universidade Estadual de Campinas, Unicamp \\ 13083-970, Campinas, S{\~a}o Paulo (Brazil).}%

\author{A. Usher}
\author{S. B. Hubbard}
\affiliation{School of Physics, University of Exeter, \\Stocker Road, Exeter EX4 4QL (UK)}

\date{\today}

\begin{abstract}
We have studied the effect of structural disorder on the de Haas van Alphen (dHvA) and Shubnikov de Haas (SdH) quantum oscillations measured in single-crystal and highly oriented pyrolytic graphite samples at temperatures down to 30 mK and at magnetic fields up to 14 T.  The measurements were performed on different samples characterized by means of x-ray diffraction, transmission electron microscopy and atomic-force microscopy techniques.  Our results reveal a correlation between the amplitude of quantum oscillations and the surface roughness. 
\end{abstract}

\maketitle

Graphite is the allotrope of carbon consisting of weakly-bonded layers of graphene.  Research in the past decade has shown that graphite exhibits some of the properties of graphene, most notably the presence of Dirac fermions \cite{Zhou_nature_2006, Luk_2009_state_of_art}, along with other remarkable properties, such as the occurrence of the quantum Hall effect (QHE), ferromagnetism  and magnetic-field-driven metal – insulator transitions \cite{PRL2006_igor_kop_QHE, PRB2002_esquin_FM_graf, kop_PRL2003_quantum_lim}.

At low temperatures, highly oriented graphite presents clear quantum oscillations in its magnetic susceptibility and electric resistivity. These effects, known respectively as the de Haas van Alphen (dHvA) and Shubnikov de Haas (SdH) effects \cite{ashcroft_book}, arise from the increase of the Landau-level degeneracy in the system as an external magnetic field (B) is increased.  They manifest as approximately sinusoidal oscillations of the material’s magnetic moment (the dHvA effect) and the electrical conductivity (the SdH effect), which are periodic in 1/B.  

It is widely known that these quantum oscillatory phenomena are suppressed by sample disorder. Textbook results show that the amplitude $\Delta$M of SdH and dHvA effects relate mainly to the width $\Gamma$ of the Landau levels, which is affected by the effective mass of charge carriers and electronic scattering rates \cite{PRL2006_igor_kop_QHE, Dingle_ov}
\begin{equation}
\Delta M \propto \frac{\lambda}{\sinh{\lambda}}\text{exp}\left( -2\pi \frac{\Gamma}{\hbar \omega_c}\right),
\label{eq_dM}
\end{equation}
with $\omega_c$ the cyclotronic frequency of the carrier and \mbox{$\lambda=2\pi^2 k_BT/\hbar \omega_c$}.  These parameters do not trivially relate to structural disorder of samples, as some structural faults might enhance electronic transport instead of hindering it (e.g. heterojunctions or tensile strains are responsible for high-mobility electron gases in semiconducting systems) \cite{PRB84-walukiewicz-heterostruct, IEEE-roughness}.

In graphite, the parameter usually held as a measurement of sample quality is its mosaic spread (v.g. ref. \cite{sacha_JoP_2013_NMR, SPI-supplies}). However, other disorder parameters can be equally important in characterizing sample quality, while not directly related to its mosaicity. For example, line-like defects and grain boundaries in highly oriented pyrolytic graphite (HOPG) can harbor ferromagnetic domains and strongly affect the sample’s electric transport properties \cite{sacha_JoP_2013_NMR, PSSB_marakova_ind_mag_G, Cervenka2009_FM_Nature}. In addition, reduced crystallite sizes in graphite are known to impair the material electronic mobility and increase the ratio between the G and D peaks observed in Raman spectroscopy measurements, while not necessarily affecting the sample x-ray pattern \cite{APL_cancado_2006_Raman}. Furthermore, small-angle rotational stacking faults in HOPG have been shown to cause interfaces between graphitic regions inside macroscopic samples, which are proposed to be responsible for the metallic-like behavior of graphite \cite{Esquin_2008_finite_size}.  

In particular, no study identifying the structural disorder parameters responsible for the suppression of quantum oscillations in graphite has been reported to date. For this reason, in this work, we correlate the parameters of the dHvA and SdH effects in different types of highly oriented graphite with their structural properties.  Our results strongly suggest that surface roughness is the fundamental disorder parameter affecting quantum oscillations in this material.

The samples used in this work were HOPG from SPI-Supplies \cite{SPI-supplies} of five different grades (designated SPI-I; SPI-II; SPI-III; GW and ZYB), Kish graphite \cite{kish_expl} and natural graphite.  All samples were characterized by x-ray diffractometry, magnetotransport and magnetic susceptibility measurements.  The transport experiments were performed in a Janis 9 T cryostat operating in the temperature range \mbox{2 K $<$ T $<$ 300 K}, using a standard 4-probe measurement technique. Magnetization experiments were performed at temperatures ranging from 30 mK to 4 K using a commercial Quantum Design SQUID magnetometer and using the torsion-balance magnetometer described in \cite{RSI_Usher}.  The magnetic field was applied parallel to the sample c-axis in the case of the SQUID and Janis-9T experiments, and at $20^o$ with respect to the c-axis in the case of the torsion-balance experiments.

\begin{figure}[h]
\includegraphics[width=6cm]{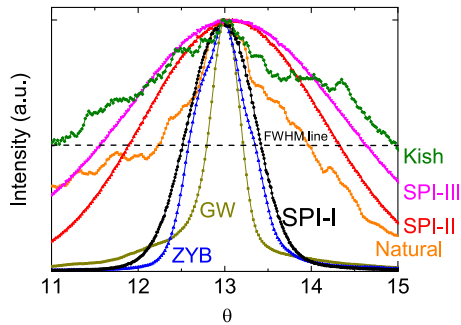}
\caption{(color online) Rocking curves for all the samples studied. The dashed line represents where the FWHM is measured. All curves were normalized to the unity.}
\label{fig_1}
\end{figure}

Initially all the HOPG samples were characterized by means of x-ray diffractometry. Rocking-curves were measured around the [0001] peak of graphite at $2\theta=26^o$ and are presented in Fig. \ref{fig_1}.  All the curves have been normalized to unity to make comparison simpler. Samples GW, ZYB, Kish and natural graphite presented curves with several peaks and will be labeled as group M. Samples SPI-I, SPI-II and SPI-III showed a single peak behavior and will be called group S. The multiple peak behavior in the M group suggests that samples in this group are composed by large, slightly misaligned, blocks of HOPG. The full-widths at half-maximum (FWHM) for each curve were extracted as shown in the figure. The values are tabled in table \ref{table_1} and are considered the mosaic spread (or mosaicity) of the samples.  

\begin{figure}[h]
\includegraphics[width=6cm]{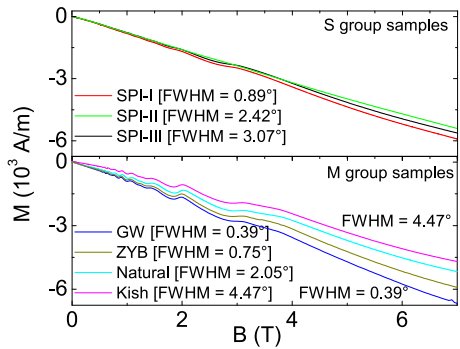}
\caption{(color online) Magnetic moment vs magnetic field at T = 2 K for different HOPG grades.  The upper panel shows the curves for S samples; the lower panel shows the curves for the M samples.  Oscillation amplitudes in the upper panel are smaller than those in the lower panel, and no correlation between susceptibility and FWHM is observed. }
\label{fig_2}
\end{figure}

\begin{figure}[h]
\includegraphics[width=6cm]{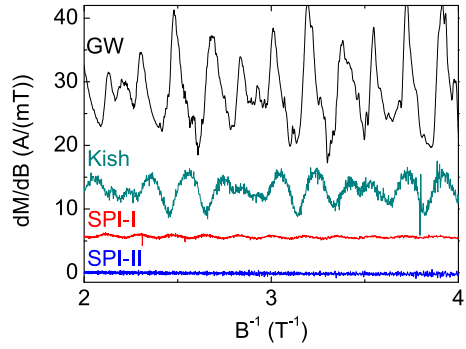}
\caption{(color online) dHvA oscillations for four different types of graphite. All measurements were carried out on a torque magnetometer at a temperature of 30 mK.  Successive curves are offset vertically for clarity.  S samples present much weaker oscillations.  In particular, SPI-II has no visible oscillations.}
\label{fig_3}
\end{figure}

Magnetization vs. magnetic field (M(B)) measurements in a SQUID magnetometer at T = 2 K showed the presence of the dHvA effect in all samples, with the exception of SPI-II. The results, presented in figure \ref{fig_2}, show that S samples have weaker quantum oscillations than M samples, regardless of their values of FWHM. This result suggests that the sample mosaicity cannot be held accountable for the amplitude of the quantum oscillations measured. For example, the samples Kish ($FWHM = 4.47^o$) and natural graphite ($FWHM = 2.05^o$) clearly present much stronger dHvA effect than SPI samples (max. $FWHM = 3.07^o$), while having a wider mosaic spread. This difference is more evident at lower temperatures, as shown in Fig. \ref{fig_3} for samples GW, Kish, SPI-I and SPI-II. 

\begin{figure}[h]
\includegraphics[width=6cm]{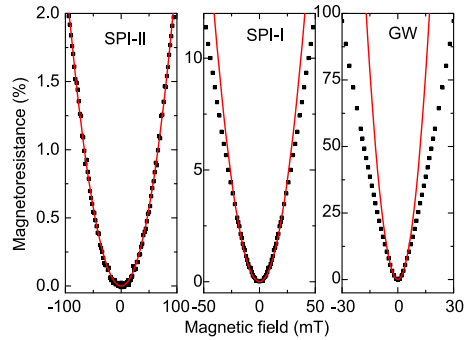}
\caption{(color online) Spectral intensity of the dHvA oscillations of four different samples for T = 30 mK.  The upper panel shows the results obtained for two of the M samples.  Lower panels are the results for the S ones.  Note the clear presence of two peaks in M samples, in contrast with only one in the S ones.  No peak was observed for SPI-II. }
\label{fig_4}
\end{figure}

\begin{figure}[h]
\includegraphics[width=6cm]{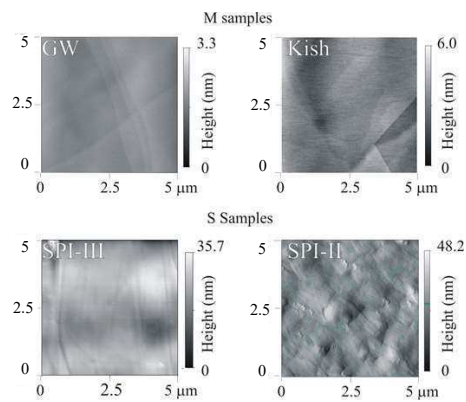}
\caption{(color online) SdH measurement for SPI-II (S) and GW (M) samples.  Despite the great disparity in oscillation amplitude between these samples, both of them show the same oscillating frequency.  The inset shows the magnification of the SPI-II data.}
\label{fig_5}
\end{figure}

We further note that samples belonging to the S group present dHvA oscillations which are not only weak, but also possess only a single frequency component, while the stronger oscillations of the M samples contain two frequency components. This is clearly seen in the Fourier transforms of the oscillating components of magnetization, as presented in Fig. \ref{fig_4}.  In this figure, the lower and higher frequencies are labeled as $\nu_1$ and $\nu_2$, respectively. The SPI-II sample had no dHvA oscillations down to 30 mK.  However, SdH measurements in this graphite have shown the presence of oscillatory resistivity with the same frequency observed for M samples, albeit with strongly suppressed amplitude (see Fig. \ref{fig_5}).  The single frequency observed for S samples coincides with the $\nu_2$ frequency observed for M samples, indicating that this carrier group has the same concentration in all graphite measured \cite{ashcroft_book}.  The values of $\nu_1$ and $\nu_2$ extracted for all samples are shown in table \ref{table_1}.

\begin{figure}[h]
\includegraphics[width=6cm]{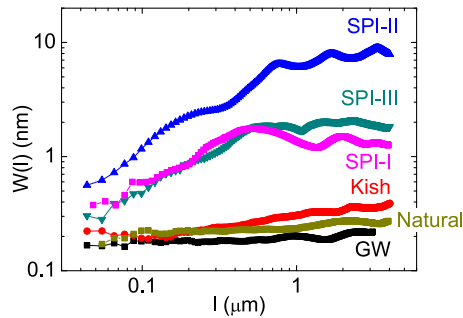}
\caption{(color online) Magnetoresistance measurements for samples SPI-II, SPI-I and GW at T = 2 K.  The points correspond to the experimental data and the lines to a fit of the low field limit by a parabola $\text{MR} = \mu^2 B^2$, with $\mu=1.36 \text{ T}^{-1}$  for SPI-II, $5.8 \text{ T}^{-1}$ for SPI-I and $117 \text{ T}^{-1}$ for GW.}
\label{fig_6}
\end{figure}

The suppression of the dHvA oscillations in the S samples can be attributed to a reduced electronic mobility on them. This is confirmed by low-field magnetoresistance (MR) measurements.  All samples presented a MR of the type
\begin{equation}
MR \equiv \frac{R(B)}{R(B=0)}-1 = \mu^2 B^2,
\label{eq_MR}
\end{equation}
which is the field dependency expected from the Drude theory for a perfectly compensated semimetal. The mobility $\mu$ of Eq. (\ref{eq_MR}) corresponds to the geometrically averaged electronic mobility of electrons (e) and holes (h) ($\mu=\sqrt{\mu_h \mu_e}$), thus only allowing for an estimation of the average mobility $\mu$ of charge carriers in different graphite. Samples with stronger quantum oscillations had higher mobilities, as illustrated for SPI-I, SPI-II and GW in Fig. \ref{fig_6}. The sample GW, which had stronger quantum oscillations among all measured graphite (see Fig. \ref{fig_9}), presented a value of $\mu=1.17\times 10^6 \text{cm}^2 \text{V}^{-1} \text{s}^{-1}$. This value is two orders of magnitude higher than what was estimated for SPI-II ($\mu=1.36\times 10^4 \text{cm}^2 \text{V}^{-1} \text{s}^{-1}$), which did not show any dHvA oscillations down to 30 mK (see fig. 3). The average electronic mobilities for all samples are presented in table \ref{table_1}. Values in brackets correspond to the uncertainty in the least significant digit of data.

\begin{table*}
\caption{\label{table_1}Summary of diffractogram, dHvA and electronic mobility results.}
\begin{ruledtabular}
\begin{tabular}{cccccccc}
 &GW&ZYB&Natural&Kish&SPI-I&SPI-II&SPI-III\\
\hline
FWHM& $0.39^o$ & $0.76^o$ & $2.05^o$ & $4.47^o$ & $0.89^o$ & $2.42^o$ & $3.07^o$\\
\thead{Rocking curve\\ peaks}&\thead{ Multiple\\ (M)} & \thead{ Multiple\\ (M)}& \thead{ Multiple\\ (M)} &\thead{ Multiple\\ (M)} & \thead{Single\\(S)} & \thead{Single\\(S)} & \thead{Single\\(S)} \\
Frequency $\nu_1$ (T) \footnotemark[1]& 4.1(1) & 4.8(3) & 4.4(1) & 4.3(1) & N/A\footnotemark[2] & N/A\footnotemark[2] & N/A\footnotemark[2] \\
Frequency $\nu_2$ (T) \footnotemark[1]& 5.6(1) & 6.5(3) & 6.1(3) & 5.8(1) & 5.9(1) & N/A\footnotemark[2] & 5.9(1) \\
\thead{Estimated average \\ mobility ($\text{cm}^2 \text{V}^{-1} \text{s}^{-1}$)}& $1.17(5) \times 10^6$ & $5.88(4)\times 10^5$ & N/M\footnotemark[3] & $1.6(4)\times 10^5$ & $5.80(6) \times 10^4$ & $1.36(5)\times 10^4$ & $6.39(5)\times 10^4$ \\
\end{tabular}
\end{ruledtabular}
\footnotetext[1]{Extracted from torque measurements at T = 30 mK.}
\footnotetext[2]{Not applicable.}
\footnotetext[3]{Not measured.}
\end{table*}

The fact that the quantum oscillations are more suppressed in samples with lower electronic mobility is not surprising. However, the question remains on the structural disorder parameters responsible for such suppression, since one cannot correlate different values of $\mu$ with the samples’ FWHM. To explore the influence of other disorder parameters on the dHvA effect in graphite, we proceed to examine the topography of our samples.  The topographies were measured by in-air Atomic Force Microscopy (AFM).  High-resolution conical tips were used in contact mode \cite{AFM_review}. 

\begin{figure}[h]
\includegraphics[width=8cm]{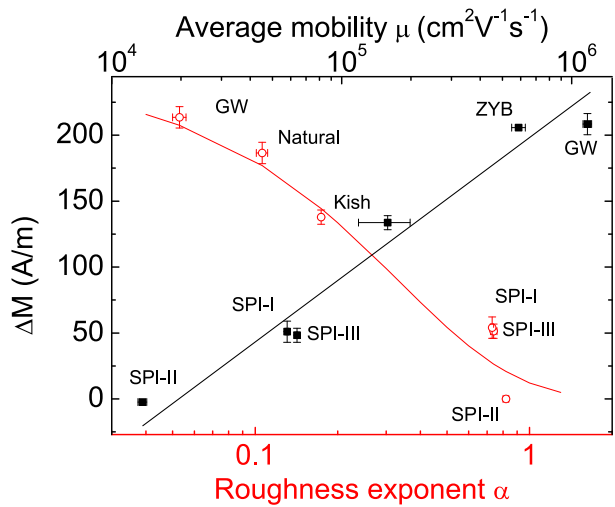}
\caption{(color online) AFM topographies for four of our samples. The lateral size of all the images is 5 $\mu$m. Note the different height scales on the right of the figures.  Samples belonging to the S group (below) show a much more corrugated surface than samples from the M group (above).}
\label{fig_7}
\end{figure}

Figure \ref{fig_7} shows the AFM images of the GW, Kish, SPI-II and SPI-III samples.  All the samples M and S have shown similar results within their groups.  The images were obtained from a freshly exposed surface and its main features did not depend on the position where they were measured.  We see that SPI-II and SPI-III (lower panels) show a similar large-scale corrugation on the surface, which cannot be observed in the morphologies of GW and Kish (upper panels).  The corrugation is associated with larger height variations in the SPI samples, as we note from the different height scales in Fig. \ref{fig_7}. From the figure, one also notices that the in-plane length of such corrugations in S samples is about 10-100 times smaller than for M samples. For example, the average lateral size of corrugations in SPI-II is approximately 100 nm, while for the GW sample this value is 2-3 $\mu$m. 

These lateral corrugation sizes are comparable with the mean free paths $l^*$ of charge carriers in our samples, estimated from the measured average electronic mobility ($\mu=el^*/(v_f m^*)$) of our graphite. Assuming the effective mass of charge carriers in graphite as \mbox{$m^*\approx 0.05 m_e$} and considering the Fermi velocity \mbox{$v_f \approx 10^6$ m/s} \cite{PR_1964_Soule}, we obtain  $l^*\approx 10$ $\mu$m for GW and $l^*\approx 100$ nm for SPI-II, which are similar to the lateral sizes of corrugations in our samples (see Fig. \ref{fig_7}) and within values observed for HOPG ($0.1 \mu\text{m}< l^* <10 \mu \text{m}$) [20, 21, 22].  This suggests that the corrugations observed in the AFM measurements act as the main source of electronic scattering in the material, limiting the electronic mobility of the sample and suppressing the quantum oscillations measured.

In order to quantify the impact of the sample corrugation on the dHvA oscillations in graphite, the morphologies of the samples were analyzed using the rms roughness:
\begin{equation}
W(l) \equiv \left[ \left\langle \left(h \left( \vec{r} \right) -\left\langle h_l \left( \vec{r}\right)\right\rangle \right)^2 \right\rangle_{\vec{r}} \right] ^{1/2},
\label{eq_RMS_rough}
\end{equation}
with $l=|\vec{r}-\vec{r'}|$, $\vec{r'}$ being the origin of the system and $\vec{r}$ an arbitrary position.  $W(l)$ is a measure of the contribution to the surface roughness due to fluctuations over a characteristic length scale $l$ \cite{fischer_int}.  For large values of $l$, $W(l)$ is expected to converge or oscillate around a finite value corresponding to the average rms roughness of the surface \cite{gogolides_int_2004}.  For low values of $l$, it can be described as  $W(l) \propto l^\alpha$, where $\alpha$ is defined as the surface roughness exponent \cite{fischer_int, gogolides_int_2004}.

\begin{figure}[h]
\includegraphics[width=6cm]{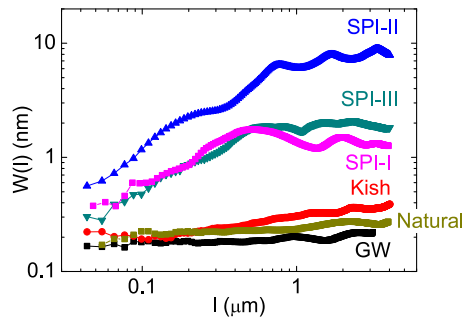}
\caption{(color online) Dependence of rms roughness on lateral length scale for six of the samples. The graph shows that the GW sample has the lowest roughness at all length scales, while SPI-II has the highest.}
\label{fig_8}
\end{figure}

\begin{figure}[h]
\includegraphics[width=6cm]{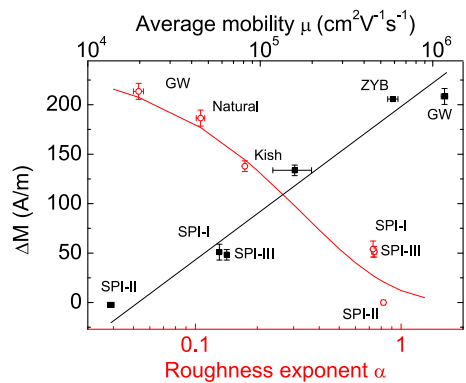}
\caption{(color online) dHvA oscillation amplitude $\Delta M$ as a function of roughness exponent $\alpha$ (open symbols, lower axis) and estimated average mobility ${\mu}$ (closed squares, upper axis) for different samples. The values of $\Delta M$ were extracted from SQUID measurements at T = 2 K and B = 2.7 T after the subtraction of a polynomial background from the M(B) curves.  The black line is a guide to the eye. The red line is a function of the type $\Delta M=0.09\times \text{exp}(-3\alpha)$.}
\label{fig_9}
\end{figure}

Figure \ref{fig_8} shows the rms roughness calculated from AFM scans ($5$ by $5$ $\mu$m in size) for all graphite measured.  Samples with corrugated morphology are associated with larger macroscopic roughness exponents (the slopes in figure \ref{fig_8}, $\alpha \approx 0.80$).  Values of $\alpha$ in the range $0.9$ - $1.0$ are expected for morphologies presenting pyramid-like features, usually called mounds in the literature \cite{gogolides_int_2004}.  In our samples, $\alpha$ values for SPI-II ($0.82$) and SPI-III ($0.73$) suggest a surface where periodic structures appear but are not so clearly defined.  Lower values of $\alpha$ are observed for the M group, $\alpha \approx 0.05$ and $0.18$ (GW and Kish).  Such values correspond to experimental observations of smoother macroscopic surfaces \cite{Krim_PRE_1993}. 

Figure \ref{fig_9} shows the dHvA oscillations amplitude $\Delta M$ of different graphite as a function of their roughness exponents $\alpha$. The values of $\Delta M$ were extracted from $M(B, T = 2 \text{K}))$ curves measured in a SQUID magnetometer. In the same figure, the electronic mobilities extracted from the magnetoresistance measurements at T = 2 K are shown.  The data clearly correlates the absence (or strong suppression) of quantum oscillations in graphite with its increased surface roughness. The dependence is roughly of the type $\Delta M \propto \text{exp}⁡(-\alpha \times \text{cte})$, which is the same functional form expected for the suppression of dHvA oscillations amplitude with the increase of sample structural disorder \cite{PRL2006_igor_kop_QHE, Dingle_ov} (see Eq. (\ref{eq_dM})). Based on these results and on our transport measurements, we conclude that the presence of wrinkles (or periodic potentials) in graphite limits the mean free path of charge carriers by being the main source of electron scattering. This results in the suppression of quantum oscillations in the samples as the surface roughness is increased.

Our results can be further analyzed in light of a theoretical model proposed Katsnelson and Geim \cite{Katsnelson_2008}, which predicts that, for weak rippling in graphene ($\alpha <0.5$), the presence of irregularities with radius $R$ and height $z$ affects the sample electronic mobility according to $\mu \propto R^2 /z^4$. Considering $R_{GW}\approx R_{Kish}$ and $z_{Kish}\approx 2z_{GW}$ (see Fig. \ref{fig_7}), this leads to an expected ratio between GW and Kish mobilities $(\mu_{GW}/\mu_{Kish})_{calc}\approx 15$. The experimental ratio (considering the measured values of $\mu$) is in the range \mbox{$6< (\mu_{GW}/\mu_{Kish})_{exp}< 10$}  (see table \ref{table_1}) and presents a wide variation due to the relatively high error associated to the measured value for $\mu_{Kish}$. Except for a factor of 3, the experimental ratio agrees with the theoretical modeling, indicating that the main source of electronic scattering in our samples with $\alpha<0.5$ is indeed the corrugation observed in AFM measurements. The same cannot be said about our samples with $\alpha>0.5$. In them, according to the model, the electronic mobility should increase with $\alpha$ following $\mu \propto n^{(2\alpha-2)}$, with $n$ the two-dimensional charge carrier concentration in the material (which is assumed to be constant in different samples based on our measured dHvA frequencies). In our SPI-samples, however, the electronic mobility $\mu$ decreases with the increase of $\alpha$ (see fig. \ref{fig_9}). This can be understood within the context of rippling if one assumes that the corrugations occurring in the S-samples are strong enough to produce resonant states, in which case the model developed by Katsnelson (for \textit{weak} rippling) is no longer applicable \cite{Katsnelson_2008}.

The reduction of electronic mobility with the increase of surface roughness in our samples is qualitatively similar to results in wrinkled graphene sheets. In graphene, the inclusion of micrometer-long folds in the sample leads to an increase of the electrical resistivity by a factor of 5-10 and a reduction of the electronic mobility up to two orders of magnitude \cite{Katsnelson_2008, Nano_xu_2009, PRL_2010_wrinkles}. Our results can also be compared to observations in high-mobility two dimensional electron gases in semiconducting heterostructures \cite{QW_Celik_1997, balkan_QW_1997}. In these systems, surface roughness is one of the key parameters limiting the electronic mobility in the junctions \cite{QW_Celik_1997}. Results in Si MOSFETS and multilayered quantum wells show that the increase of roughness in these structures increases the scattering rate of the electron gas, prompting a reduction of the amplitude of the quantum oscillations \cite{QW_Celik_1997, balkan_QW_1997, cao_APL-2007, Sakaki_APL_1987}. For example, the reduction of the lateral size of irregularities in GaAs-GaAlAs quantum wells by a factor of 3 results in a 20-times increase of the quantum oscillations amplitude in the system \cite{QW_Celik_1997}.


%

In conclusion, we have shown that sample roughness plays an important role in the suppression of quantum oscillations in graphite.  This type of disorder, though important, is not explored in most experiments concerning HOPG.  It was observed solely in AFM measurements, being concealed in x-ray diffractometry and Raman spectroscopy.   Our experiments have shown that this kind of defect does not affect the FWHM of the x-ray diffraction rocking-curves but drastically reduces the mobility of the material, and is an important disorder parameter that has hitherto been neglected.  
\section*{Acknowledgments}
We would like to thank Prof. Dr. Pablo Esquinazi and Prof. Dr. Walter Escoffier for stimulating discussions. This work was carried out with the support of CNPq (Conselho Nacional de Desenvolvimento Científico e Tecnológico - Brasil) and FAPESP (Funda\c{c}{\~a}o de Amparo a Pesquisa do Estado de S{\~a}o Paulo).


\begin{thebibliography}{29}
\bibitem[{kis()}]{kish_expl}
\bibinfo{howpublished}{Kish graphite was originally a by-product of steel
  manufacture, and it resembles natural graphite.}
\end{thebibliography}
\end{document}